   \documentclass[12pt,preprint]{aastex}

\usepackage{bm}

\begin{document}

\title{%
    How to Find More Supernovae with Less Work:
    Object Classification Techniques for Difference Imaging
}

\author{%
   S.~Bailey,\altaffilmark{1,4}
   C.~Aragon,\altaffilmark{1}
   R.~Romano,\altaffilmark{1,2}
   R.~C.~Thomas,\altaffilmark{1}
   B.~A.~Weaver\altaffilmark{1,3}
   D.~Wong\altaffilmark{1}
}

\altaffiltext{1}{Lawrence Berkeley National
Laboratory, 1 Cyclotron Road, Berkeley, CA 94720}
\altaffiltext{2}{Luis W. Alvarez Fellow, National Energy Research Scientific Computing Center, 1 Cyclotron Road, Berkeley, CA 94720} 
\altaffiltext{3}{University of California, Space Sciences Laboratory,
Berkeley, CA 94720}
\altaffiltext{4}{Corresponding author: sjbailey@lbl.gov}

\begin{abstract}
We present the results of applying new object classification techniques
to difference images in the context of the Nearby Supernova Factory
supernova search.
Most current supernova searches subtract reference images from new images,
identify objects in these difference images, and apply simple threshold cuts on
parameters such as statistical significance, shape, and motion
to reject objects such as cosmic rays, asteroids, and subtraction
artifacts.  
Although most static objects subtract cleanly, even a very low false
positive detection rate can lead to hundreds of
non-supernova candidates which
must be vetted by human inspection before triggering additional followup.
In comparison to simple threshold cuts, more sophisticated methods such as
Boosted Decision Trees, Random Forests, and Support Vector Machines
provide dramatically better object discrimination.
At the Nearby Supernova Factory, we reduced the number of non-supernova
candidates by a factor of 10 while increasing our supernova identification
efficiency.  Methods such as these will be crucial for maintaining
a reasonable false positive rate in the automated transient alert
pipelines of upcoming projects such as PanSTARRS and LSST.
\end{abstract}


\section{Introduction}

Future large scale survey projects such as
PanSTARRS\footnote{http://pan-starrs.ifa.hawaii.edu}
and LSST\footnote{http://lsst.org} are expected
to generate automated rapid turnaround transient alerts for objects
such as supernovae, active galactic nuclei, asteroids, Kuiper belt objects,
and variable stars.
They will do this by comparing new images to coadded stacks of reference
images taken previously.  Repeat observations of the same field will
occur over timescales of minutes, hours, days, months, and years.
Robust rejection of spurious non-astrophysical objects
will be crucial to avoid excessive false positive alerts.

A major difficulty of current optical transient programs is the huge
number of false positive objects which are difficult to reject while
maintaining high selection
efficiency for the real objects of interest.  For example, the 2005 Sloan
Digital Sky Survey II (SDSS-II) supernova program \citep{sdss:becker}
required objects to be detected within 0.6 arcsec in at least two filters
with signal-to-noise greater than 3, yet this generated $\sim$4,000 objects
per night which needed to be visually checked by humans for verification.
Their 2006 search drastically reduced this scanning load by requiring that
all but the brightest objects be identified at the same location
on multiple nights before they are passed to a human for verification.
Although this reduced their scanning load, this method is not
applicable to the real-time transient alert pipelines of PanSTARRS and
LSST.  A ``60-second transient alert'' would be meaningless if it really
meant ``$N$ days plus 60 seconds after the first positive identification.''
Although PanSTARRS and LSST will have multiple exposures of a field
in the same night, this is equivalent to the multiple-filter requirement
of the SDSS 2005 program which was still swamped by false positives.

This problem of false positives
is not unique to nearby transient searches; it arises whenever a large
number of objects are imaged, either from a wide-field survey or a
deep narrow survey.  The ESSENCE \citep{essence}
and SNLS \citep{snls} Canadian pipeline supernova
searches both result in 100--200 objects to
scan per night of data\footnote{private communication with
W.M. Wood-Vasey (ESSENCE) and D. Balam (SNLS)}.
Although this is a manageable load for a current experiment,
it would not scale to future surveys which will image thousands
of square degrees per night.\footnote{For comparison,
    the SNLS supernova survey covers $\sim1$ square degree per night,
    SDSS covers 150 square degrees per night,
    and SNfactory covers 350 to 850 square degrees per night.}
If current methods were used, the projects would need to drastically
reduce their signal efficiency in order to maintain a manageable
false-positive rate.
The SNLS French pipeline uses multi-night data and an artificial
neural net to select candidates for verification, but as noted above,
using multi-night information is not applicable to rapid turnaround
transient alert pipelines which intend to produce alerts within
a minute of the first positive detection.

False positives arise from a variety of sources including diffraction spikes,
saturated stars, optical ghosts, star halos, cosmic rays, satellite trails,
CCD amplifier glow,
other CCD artifacts, and image processing artifacts.  In principle all of these
effects are best identified and either fixed or masked at the image level.
In practice there will always be effects which produce spurious detections.
This problem is especially bad at the start of a search when covering new
areas of sky, before consistent problems can be identified and masked.
The goal of a classifier is to identify the real
candidates of interest (signal events) while rejecting the spurious objects
(background events).

In some cases, real astrophysical variable objects are the background events
for other analyses.  For example, asteroids, variable stars,
and active galactic nuclei form a background for nearby
supernova searches, yet they are the core science for other programs.
This paper is written from the context of a nearby supernova search
and thus these other
astrophysical events are treated as background to reject,
but the methods presented here are generally applicable to
many object classification problems.

This paper presents the results of applying modern machine learning
techniques to the supernova search pipeline of the
Nearby Supernova Factory \citep{snfactory}.
\S \ref{sec:methods} presents a variety of machine learning techniques.
\S \ref{sec:SNfactory} describes the Nearby Supernova Factory search,
and \S\S~\ref{sec:training} and \ref{sec:software} present
the training data and classification software used.
\S \ref{sec:comparison} compares the various methods.
We find that methods such as
Boosted Trees, Random Forests, and Support Vector Machines perform
dramatically better than the threshold cuts which are typically used
by supernova search programs.

\section{Classification Methods}
\label{sec:methods}

Classification methods identify signal {\it vs.}~background
events based upon a set of features (also called variables, attributes,
or scores)
which describe the events.  For example, objects in photometric images
can be described by their magnitude, signal-to-noise, and shape parameters
such as width and ellipticity.  These features can be used to distinguish
stars from galaxies, cosmic rays, or imaging artifacts.

The optimum separation of two classes of events is application dependent,
depending upon the desired tradeoff between purity (the fraction of
selected events which are real signal), completeness (the fraction of
real signal events which are selected), and the total sample size selected.
For example, a measurement
which depends upon a statistical fit to both signal and background events
might optimize the signal-to-noise ratio $\sim S/\sqrt{S+B}$
where $S$ and $B$ are the number
of real signal and background events which the classifier selects for the fit.
A supernova search algorithm, on the other hand,
might maximize the purity with the constraint that the completeness
remain above 90\%.

Some methods, such as threshold cuts (\S \ref{sec:cuts}),
produce a boolean signal/background
decision and the cuts themselves must be adjusted to optimize the separation.
Other methods have an automated training procedure and
produce a single statistic which rates how signal-like or
background-like a new event is.  The user may then cut on that statistic
to optimize the desired figure of merit.

Classifier parameters are tuned using a training dataset of known
signal and background events to optimize
the separation power.  Since the results are influenced by the particular
statistical fluctuations of the training dataset, the separation power
on the training data itself cannot be used as
a fair measure of the power of a classifier.  Instead, a separate validation
set is used to assess the performance.  If enough training data is available,
one uses one dataset to train a variety of classifiers with
different parameters, a second dataset to select which set of parameters
produces the best classifier, and a third dataset to validate the
final performance.
Ideally one trains and validates
using real data; in practice simulated data are often used for training
and validation before applying the classifier to real data.
It is important to note that the quality and power of any classifier
will be affected by the accuracy of the training sample.  One must be
careful to minimize and measure any biases introduced through a simulated
training sample which does not completely reflect real data.

\subsection{Threshold Cuts}
\label{sec:cuts}

Automated supernova searches have typically operated by applying
simple threshold cuts to the features describing objects.
For example, a supernova search might
keep objects which have a signal-to-noise ratio $S/N > 5$,
astrometric positions that agree to within 1 arcsec on 2 or more images,
and a width consistent with stars on the images ({\it e.g.}, within a
factor of 2 of the median width of stars).
If an object fails any of these cuts, it is rejected.
These cuts are easy to understand but do not reflect
the subtleties of a multidimensional space.
An object which just barely
fails one of the cuts is still rejected the same as an object which
fails many cuts.  It also does not naturally handle correlations between
the variables, {\it e.g.,} between the $S/N$ and the astrometric
accuracy.\footnote{In a simple case such as this, one could combine $S/N$ and
    astrometric positions to form an uncorrelated variable; accomplishing this
    in the general case for a large number of variables is non-trivial.}
To use threshold cuts, one must find uncorrelated variables without
significant outliers such that every cut maintains a high signal efficiency
while rejecting background.

Compared to curved boundary selections, threshold cuts are also
an inefficient way to select a subset of a hyperspace as the number
of dimensions grows large \citep{koeppen}, even for dimensions as few as 5.
{\it e.g.}, for 3 dimensions, the volume ratio of a cube to its embedded
sphere is 1.9; for 5 dimensions it is 6.1, and for 10 dimensions
it is 401.5.
This ratio goes to infinity as the number of dimensions increases.
Thus if
a set of signal events is distributed as an ellipsoid in some feature
space, an ellipsoid shaped selection contains much less volume
(and thus likely much less background) than the equivalently dimensioned
hypercube.

Although commonly used in supernova searches,
threshold cuts are widely recognized
as being a non-optimal method for signal/background separation problems.
The following sections describe a variety of more powerful techniques for
identifying supernovae in difference images.

\subsection{Multi-dimensional Probability Measures}

A more sophisticated approach models the probability distribution
function (PDF) for the signal and the background for each of the 
features.  The combined probability of all of the feature values
for an object is used to make the signal/background decision.  This
improves over threshold cuts by eliminating rejections based upon a
slightly marginal value of a single feature, but it requires a detailed
modeling of the PDF of each feature, including all correlations and
outliers in the distributions.  This suffers from
the ``curse of dimensionality'' \citep{bellman}: since the volume
of a hyperspace grows exponentially with the number of dimensions,
the size of a training sample must also grow exponentially to
adequately determine the PDFs.

\begin{figure}
\centering
\includegraphics{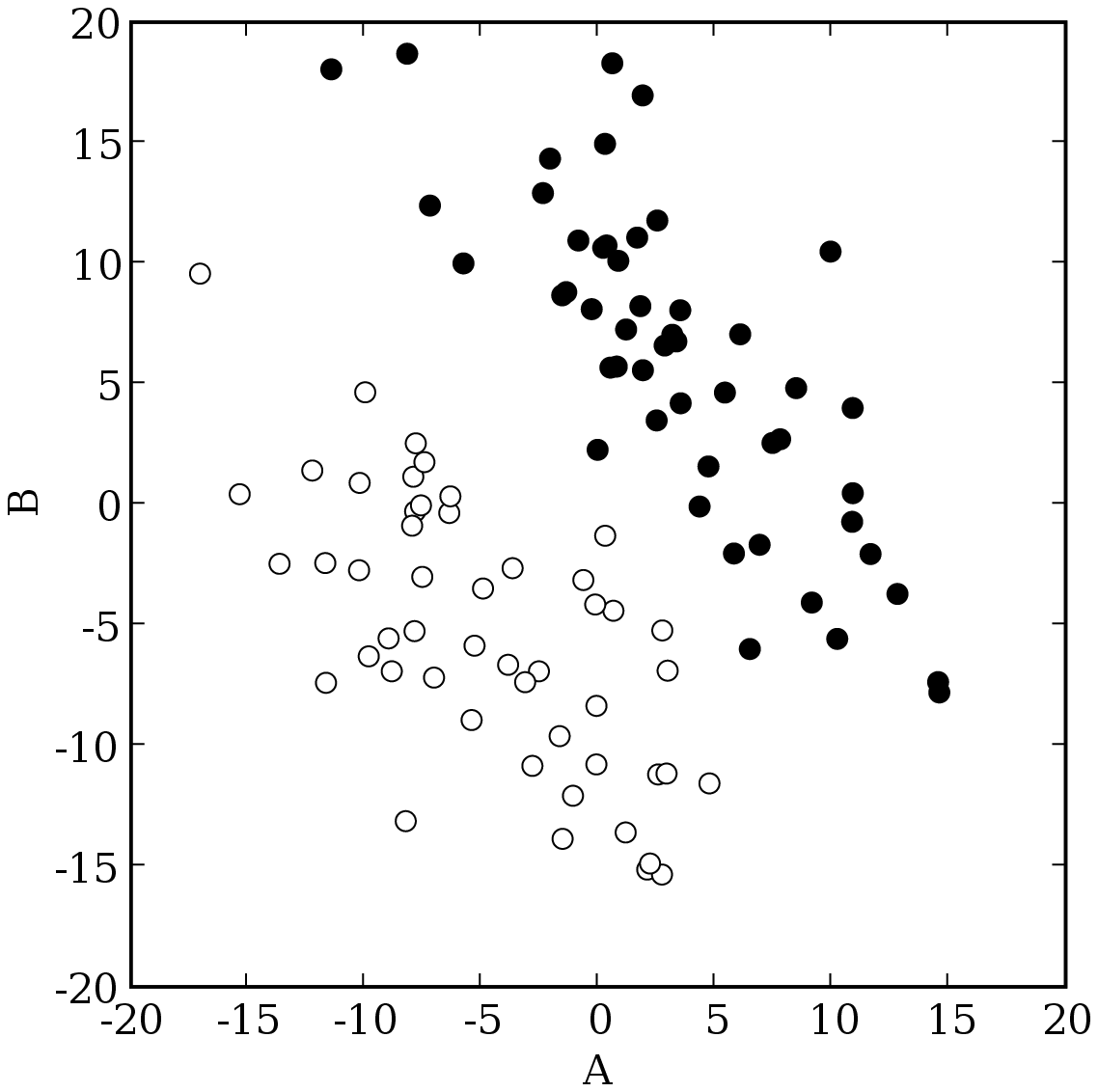}
\figcaption{
Example data which would be well separated using Fisher
Discriminant Analysis.  The two classes of events (open and filled
circles) are not well separated by either feature $A$
or $B$, but their correlation is such that the combination
$A + B$ provides very good separation of the two classes.
\label{fig:fisher}
}
\end{figure}

If the signal and background features are Gaussian distributed with
only linear correlations, Fisher Discriminant Analysis \citep{fisher}
finds the best linear combination of
features to maximize the separation of the two classes.
Figure \ref{fig:fisher} shows a toy example of
data which would be well separated using Fisher
Discriminant Analysis.  The two classes of events (blue triangles
and red squares) are not well separated by either feature $A$
or $B$, but their correlation is such that the combination
$A + B$ provides very good separation of the two classes.

More generally, if a set of events $\{{\bf x}\}$ in some feature space
have means ${\bm \mu}_{0,1}$ and covariances
$\Sigma_{0,1}$ for classes 0 and 1, then a linear combination
${\bf w} \cdot {\bf x}$
will have means ${\bf w} \cdot {\bm \mu}_{0,1}$ and covariances
${\bf w}^T \Sigma_{0,1} {\bf w}$, where ${\bf w}$ is a set of
coefficients defining a linear combination of the features.
The separation of the two classes may be defined as
\begin{equation}
\Delta = {({\bf w} \cdot {\bm \mu}_0 - {\bf w} \cdot {\bm \mu}_1)^2 \over 
    {\bf w}^T \Sigma_0 {\bf w} + {\bf w}^T \Sigma_1 {\bf w} },
\end{equation}
{\it i.e.}, the separation of the means is measured in units of the
variances.
Fisher showed that the maximum separation is achieved when
\begin{equation}
{\bf w} = (\Sigma_0 + \Sigma_1)^{-1} ({\bm \mu}_1 - {\bm \mu}_0)
\end{equation}
The means and covariances of the signal (1) and background (0) classes
may be estimated from a training sample, and thus the calculation of
the best linear combination for separating the classes
is simply a matrix inversion.  This method
breaks down when there are non-linear correlations or when there are
significant outliers or otherwise non-Gaussian variances such that a
simple mean
and covariance is not a good descriptor of the feature distributions
for the two classes.  In practice, Fisher Discriminant Analysis is
most often used to combine several linearly correlated features
into a single feature to reduce the dimensionality of a problem before
applying another classification method.
    
\subsection{Decision Trees}

\begin{figure}
\centering
\includegraphics{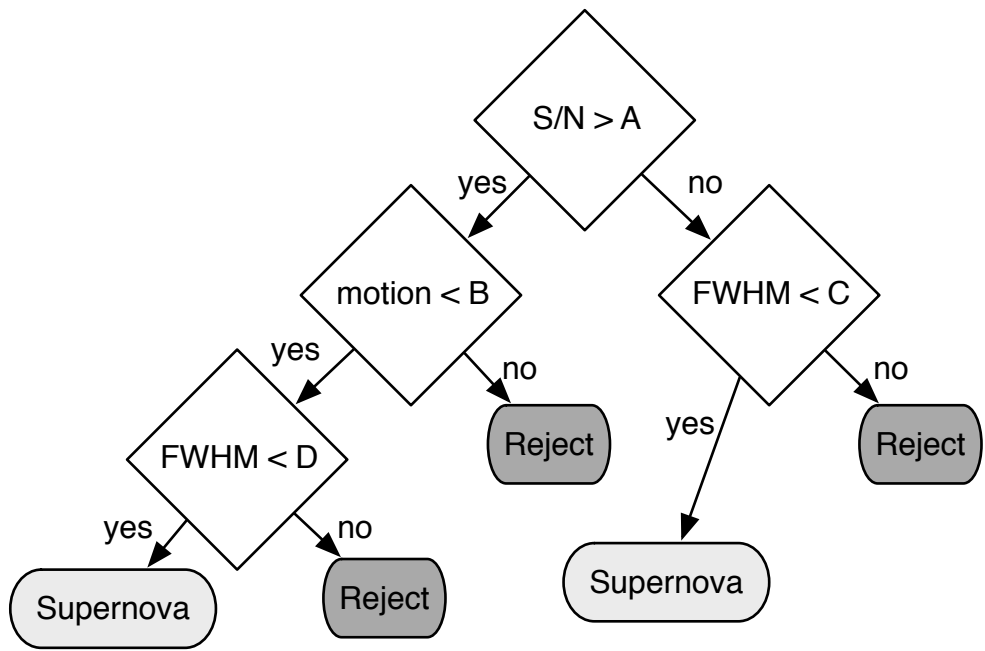}
\figcaption{
Example decision tree which would treat high signal-to-noise objects
differently than low signal-to-noise objects.  In practice, a real
decision tree has many more branches and the same variable can be
used to branch at many different locations with different cut values.
\label{fig:decisiontree}
}
\end{figure}

Decision trees \citep{breiman}
separate signal from background events by making a
cascading set of event splits as shown in Figure \ref{fig:decisiontree}.
This forms a generalization of threshold cuts by
selecting many hypercubes in the multi-dimensional feature space
rather than a single hypercube of cuts.  The training procedure
described below automatically selects the features and cut values
to generate a tree with maximal separation of signal and background
events.

The training procedure begins with a sample of training events and
considers all features and cut values to form
two subsets with the best separation of signal and background.
The procedure is recursively applied to each of the subsets to form
further branches.  The recursion is stopped when some condition is
met, {\it e.g.}, the subset is entirely signal or background, or
the subset has reached a minimum allowed size
(a minimum size requirement prevents
overtraining on statistical fluctuations of small samples).
The terminal nodes which are not further split are called leaves,
and are assigned as either signal or background leaves depending
upon the training events which ended up on those leaves.

There are a variety of ways to define the best separation at each
split; for this study we used the Gini parameter \citep{gini, breiman},
which is widely used and provides robust performance.
Define the purity of a sample of training events as
\begin{equation}
P = {\sum_S w_S \over \sum_S w_S + \sum_B w_B}
\end{equation}
where the sums are over the signal events $S$ and background events $B$
and $w_i$ are a set of event weights.  Typically all of the weights
are the same and their absolute normalization is arbitrary.
If needed, relative weights may be used to increase
the influence of an underrepresented subsample of the training data.
The role of weights will be more important in the Boosted Trees
method described in \S \ref{sec:boostedtrees}.
Note that $P=1$ for a sample of pure signal events, $P=0$ for a sample of pure
background events, and $P(1-P)=0$ for a sample which is either purely
signal or purely background.

Define
\begin{equation}
{\rm Gini} = P(1-P) \sum_{i=1}^n w_i 
\end{equation}
where the sum is over all events in that sample.
At each node, the training procedure considers all possible
features and cut values to minimize the quantity
\begin{equation}
{\rm Gini}_{\rm left\ child} + 
{\rm Gini}_{\rm right\ child}
\end{equation}
to find the best separation of events.
If this split would not increase
the overall quality of the tree, {\it i.e.},
\begin{equation}
{\rm Gini}_{\rm parent} <
{\rm Gini}_{\rm left\ child} + 
{\rm Gini}_{\rm right\ child}
\end{equation}
then the node is left as a leaf node, assigning it as a signal leaf
if $P>0.5$ and a background leaf otherwise.  If the split would increase
the overall quality of the tree, the events are split into two nodes
and the procedure is recursively applied to each of those nodes until
the stopping conditions are met ({\it e.g.}, minimum leaf sizes) or no
splits can be found which would improve the overall quality of the tree.

Decision trees are a generalization of threshold cuts and thus have more
flexibility to optimally select a set of signal events within a feature space.
However, single decision trees tend to be unstably dependent upon
the details of the
training set.  A small change in the training set can produce a
considerably different tree and thus a considerably different performance
on the validation set.

\subsubsection{Boosted Trees}
\label{sec:boostedtrees}

Boosting algorithms improve the performance of a classifier by
giving greater weight to events which are hardest to classify.  In the
case of decision trees, a tree is trained on a set of data,
misclassified events are identified and their weights are increased, and the
process is repeated to form new trees.  This iteratively produces
a set of increasing quality decision trees.
The final classifier uses the weighted ensemble average
of all of the trees to make a classification decision.  The boosting
provides decision trees with better separation power, and the ensemble
average washes out the training instabilities associated with single
decision trees.  In applications with $\sim$20 or more input features,
Boosted Decision Trees can provide significantly better results than
Artificial Neural Networks \citep{miniboone}; see also \S \ref{sec:ann}.

There are a variety of boosting algorithms used to increase the weights
of misclassified events
\citep{bdt:freund96, bdt:friedman01, bdt:friedman00}.
We describe here the commonly used
Discrete AdaBoost method \citep{bdt:freund96}.
Define the error rate for tree $m$ as
\begin{equation}
{\rm err}_m = {\sum_{i=1}^{N} w_i I_i \over \sum_{i=1}^N w_i}
\end{equation}
where $I_i = 0$ if event $i$ is correctly classified and $I_i = 1$
if it is incorrectly classified.  Typically the first tree is trained
with the same weight for all events.
Then adjust each of the event weights using
\begin{eqnarray}
\alpha_m    & = & \beta \times \ln[ (1-{\rm err}_m) / {\rm err}_m ]     \\
w_i & \to & w_i \times e^{\alpha_m I_i}
\end{eqnarray}
This increases the weights of misclassified events; the weights are
increased more when the tree has a low error rate.
These new weights are then used to generate a new decision tree.
The standard AdaBoost algorithm uses $\beta=1$ but this can be adjusted to
vary how quickly the weights are updated with each iteration.

After generating $M$ individual trees with weights $\alpha_m$,
the final classifier answer for an event described by
a set of features ${\bf x}$ is
\begin{equation}
T({\bf x}) = \sum_{m=1}^M \alpha_m T_m({\bf x})
\end{equation}
where $T_m({\bf x})$ is the result for tree $m$: 
0 if ${\bf x}$ lands
on a background leaf and +1 for a signal leaf.
The absolute normalization of $T({\bf x})$ is arbitrary; we
chose to renormalize the $\alpha_m$ weights such that
$0 \le T({\bf x}) \le 1$.

\subsubsection{Random Forests}

Random Forests \citep{rf:breiman}
also generate multiple decision trees for a given
training set and use a weighted average of the trees as the final
decision metric.  When training a tree, at each branch the
training cycle only considers a random subset of the possible features
available to use.  This has the effect of washing out
the typical training instabilities of decision trees and produces
a classifier which is fast to train and robust against outliers.

\subsection{Support Vector Machines}

The Support Vector Machine (SVM) algorithm is a classification
method that has successfully been
applied to many pattern recognition problems and is founded on
principles of statistical learning theory~\citep{svm:va98, svm:chen05}.
It nonlinearly
maps data points from the original input space to a higher-dimensional
feature space in which an optimal hyperplane parameterized by a normal
vector ${\bf w}$ and offset $b$ is computed such that the separation between
events in different classes is maximized. The linear decision boundary
is defined as $ f({\bf x}) = {\bf w} \cdot {\bf \phi}({\bf x}) + b$,
where ${\bf x}$ is a vector in the feature space which describes objects and
${\bf \phi}$ is a mapping which embeds the problem into a
higher-dimensional space in which classes are more easily
separable than in the original feature space.

An optimization problem is
constructed to find the unknown hyperplane parameters, and
the optimal hyperplane normal ${\bf w}$ is found to be entirely
determined by the subset of events nearest to the optimal
decision boundary (also called support vectors, ${\bf x}_i$) as
follows: 
${\bf w} = \sum_i c_i \phi({\bf x}_i)$,
where the coefficients $c_i$ are 
the Lagrange multipliers used in solving the nonlinear
optimization and are a byproduct of the optimization.


The hyperplane parameters are solved by maximizing the margin
(the distance between the hyperplane and the example events in each class),
which is formulated as a nonlinear
constrained optimization problem, where the constraints
enforce that examples from different classes lie on opposite
sides of the hyperplane.
The objective function to be minimized is convex, {\it i.e.},
it is guaranteed to have a global minimum and no local minima.
The linear decision boundary corresponds to a nonlinear
(and possibly disjoint)
decision boundary in the original feature space. Once the hyperplane is
found, a set of features ${\bf x}$ is typically classified into one of
the two classes by applying a threshold cut to $f({\bf x})$.

Rather than calculating $\phi({\bf x})$ explicitly while
evaluating
${\bf w} \cdot \phi({\bf x}) =
\sum_i c_i \phi({\bf x}_i) \cdot \phi({\bf x})$,
the actual embedding is
achieved through a kernel function defining an inner product in the
embedding space,
$k({\bf x}_1,{\bf x}_2) = {\bf \phi}({\bf x}_1) \cdot
{\bf \phi}({\bf x}_2)$.
This ``kernel trick'' makes class prediction easy to implement
and fast to compute.  Several common kernel mappings are
given in \cite{svm:chen05}.
In practice, the kernel function is typically chosen empirically
via training and testing, and the simplest function giving the
desired performance is used.  The Gaussian kernel used in this analysis
\begin{equation}
k({\bf x}_1,{\bf x}_2) = \exp( -||{\bf x}_1 - {\bf x}_2||^2 / 2 \sigma^2)
\end{equation}
is commonly used because it only has one free parameter to be
tuned ($\sigma$)
and empirically performs as well as, if not better than, more
complex kernels which may overfit the data.


For this analysis we used a soft-boundary SVM method
called $C$-SVM, which handles noisy data
with high class overlap by adding a regularization term to the objective
function.  This term allows but penalizes training
points lying on the wrong side
of the decision boundary. The regularization parameter, $C$,
controls the trade-off between maximizing the separation
and allowing some amount of training error while finding the
hyperplane which maximally separates signal from background.

The advantages of SVMs include the existence of a unique solution, the
simple geometric interpretation of the margin maximization function, the
capacity to compute arbitrary nonlinear decision boundaries while
controlling over-fitting with soft margins, the low number of parameters
to be tuned (as few as two, depending on the choice of kernel), and the
dependence of the solution on only a small number of data points
(the support vectors)
which define the boundary of the class separation hypersurface.
For SVM implementation details, see \cite{svm:va98}.

\subsection{Artificial Neural Networks}
\label{sec:ann}

Artificial Neural Networks (ANNs) are a broad category of
classification methods
originally inspired by the interconnected structure of neurons
and synapses in the brain.
These methods map a set of input variables to one or more output results
via one or more ``hidden layers'' of intermediate nodes.
For a supernova search, the inputs would be the features describing
each object and the desired output would be 1(0) for signal~(background).
For an overview of these methods, see \cite{ann:bishop}.


ANNs can be powerful classifiers and have been used in many applications,
though they are slow to train and require some experimentation to
optimize the number of hidden layers and nodes to match a given problem.
They also do not scale well with an increasing number of input
features, and their results become unstable when there are significant
outliers or otherwise irrelevant input data.
For these reasons,
ANNs were not deemed to be an appropriate classification method for
our dataset and this method was not
pursued for this study.

\section{Nearby Supernova Factory Search}
\label{sec:SNfactory}

The Nearby Supernova Factory \citep{snfactory} search uses
data from the Near Earth
Asteroid Tracking (NEAT) program\footnote{http://neat.jpl.nasa.gov}
and the Palomar QUEST consortium\footnote{http://hepwww.physics.yale.edu/quest/palomar.html}
using the 112 CCD QUEST-II camera \citep{questcamera}
on the Palomar Oschin 1.2-m telescope.
The NEAT observing pattern obtains triplets of 60-second exposures spread
over a time period of $\sim$1 hour using a single RG610 filter,
which is a long pass filter redward of 610 nm.
This allows the search to distinguish between asteroids, whose motion is
typically detectable
on that timescale, and spatially static objects such as supernovae.
The QUEST data are obtained in 4 filters in driftscan mode; our search
uses the two filters
which cover the best quality CCDs
(either Bessel $R$ and $I$ or Gunn $r$ and $i$
depending upon camera configuration).
The QUEST data cover less area and tend to be
cosmetically cleaner than the NEAT data, resulting in fewer spurious
detections overall.
Since the false positive background
events are much worse in the NEAT data, our study of alternative
classification methods has focused on the NEAT dataset.

Coadded stacks of images taken from 2000 to 2003 are used as references.
The new and reference
images are convolved to match their point-spread-functions (PSFs),
the fluxes are normalized by
matching stars, and the reference is subtracted from the new
images.  Objects in the subtraction are identified based upon contiguous
pixels with $S/N > 3$ with at least one pixel with $S/N > 5$.
Objects are described by features such as position,
full-width-half-max (FWHM) in $x$ and $y$,
aperture photometry and associated uncertainties in 3 apertures,
distance to nearest object
in the reference coadd, and measures of the roundness and irregularity
of the object contour
based upon Fourier descriptors \citep{zahn}.
Additional features are formed as combinations of features from
the same object observed on multiple images.
Combined features include
the object motion between two images and the consistency of the statistical
significance of the measurements in different images.  The features are
used by a classification method (originally threshold cuts, more recently
Boosted Decision Trees)
to select supernova candidates of interest which are then visually
scanned by humans to select the best candidates for spectroscopic
confirmation and followup by the SuperNova Integral Field Spectrometer
(SNIFS) \citep{snfactory}, on the University of Hawaii 2.2-m telescope on
Mauna Kea.

\section{Training Dataset}
\label{sec:training}

To generate signal events for training, fake supernovae were
introduced into the images by moving real stars of a desired magnitude
to locations distributed about known galaxies on the SNfactory search
images.  By using real stars from the same images as the galaxies,
we realistically model the point-spread-function, noise, and possible
image artifacts present in that image.  These images with fake supernovae
were processed with the same data analysis pipeline as real images
to identify objects and measure their features for classification.

The stars are sampled in a circular region of 20 pixels in diameter.
The typical FWHM of stars on these images is about 3 pixels,
so this samples the PSF out to $\sim8\sigma$.
The average sky level of the image is subtracted
from the sampled pixels before they are added in the new location, which
implicitely assumes that the sky level is uniform over the image.
This is valid in
most cases, and it is simple to reject fakes created from cases that
violate this assumption.  Typically these fakes will have a FWHM that
differs very significantly from the average ($\sim$20 pixels vs. $\sim$3).
Most stars are sufficiently isolated that they do not bring along portions of
other objects, and we identify and reject cases where this does
happen.  The spatial variation of the PSF is minimal compared to the
night-to-night variations which much be addressed by the image subtraction
pipeline, thus this fake supernova generation procedure does not attempt
to correct for the small spatial variations of the PSF across the CCD.

Background events were randomly selected from 4.1 million
other objects identified on the subtractions with fake supernovae.
These subtractions covered a month of data taking including bright
and dark times and a variety of seeing conditions.
Objects within 20 pixels of a fake supernova were excluded to
avoid any artifacts which might be introduced through an ill-formed fake.
These background events form a randomly selected subset of the genuine
backgrounds faced by the supernova search in the real data, and thus
represent the real fractions of each type of background event faced.

The signal and background samples were split into training and validation
subsets.  Several training sets were formed with 5,000 signal and 5,000
background events each.  The final validation was performed using 20,000
signal and 200,000 background events.  The training dataset for the
Support Vector Machine method was augmented with real supernova
discoveries in an attempt to improve its overall performance.
The original training set used 19 features; an additional 13 features
were then added which improved the performance of the Boosted Trees
and Random Forests but decreased the performance of the SVM.
The results shown in \S \ref{sec:comparison} are for the best performance
achieved for each classifier ({\it i.e.}, using 19 features for SVM
        and 32 features for the other methods).

\section{Classification software}
\label{sec:software}

For Fisher Discriminant Analysis, Boosted Trees, and Random Forests,
we used the open-source C++ software
package StatPatternRecognition.\footnote{
http://sourceforge.net/projects/statpatrec}
Training a set of 200 boosted trees
using 10,000 training events with 19
features each with a minimum leaf size of 15
events\footnote{The selection of the number of trees and minimum leaf
    size are described in section \ref{sec:paramchoice}.}
takes $\sim$3 minutes (wallclock) on a 2 GHz AMD Opteron CPU.
Training with 32 features takes $\sim$45 minutes.
Once the trees are trained, it takes approximately 0.6 ms (wallclock)
to evaluate the results for an object.
Random Forests took less than 2 minutes to train on the dataset
with 32 features, using the same parameters as above.  Evaluation
of a new event takes approximately 0.2 ms.
Fisher Discriminant analysis took 1 second to train on the
32 feature dataset and 0.07 ms to evaluate results for a new object.

For SVM, we used the LIBSVM C++ package.\footnote{
http://www.csie.ntu.edu.tw/$\sim$cjlin/libsvm}
Training a $C$-SVM
using 10,000 training events with 19
features each takes from 5 to 15 seconds (wallclock)
on a 2 GHz AMD Opteron CPU,
depending on the settings of the two parameters used (if the parameters
overfit the data, more support vectors are needed so training time
increases).  Evaluating the SVM on a new data point takes approximately
0.6 ms (wallclock).

\section{Comparison of Methods}
\label{sec:comparison}

In the end, most classification methods produce a single classification
statistic with arbitrary normalization which
rates how signal-like or how background-like the candidate is.
A threshold cut on this statistic can be used to select a subsample of
events with desired signal {\it vs.}~background purity.
A useful way to visualize the power of a classifier is
to plot the fraction of false positives ({\it i.e.}, background
events incorrectly classified as signal) {\it vs}.~fraction of true positives
({\it i.e.}, signal events correctly identified)
for various selection values on the classification statistic.

\begin{figure}
\centering
\includegraphics{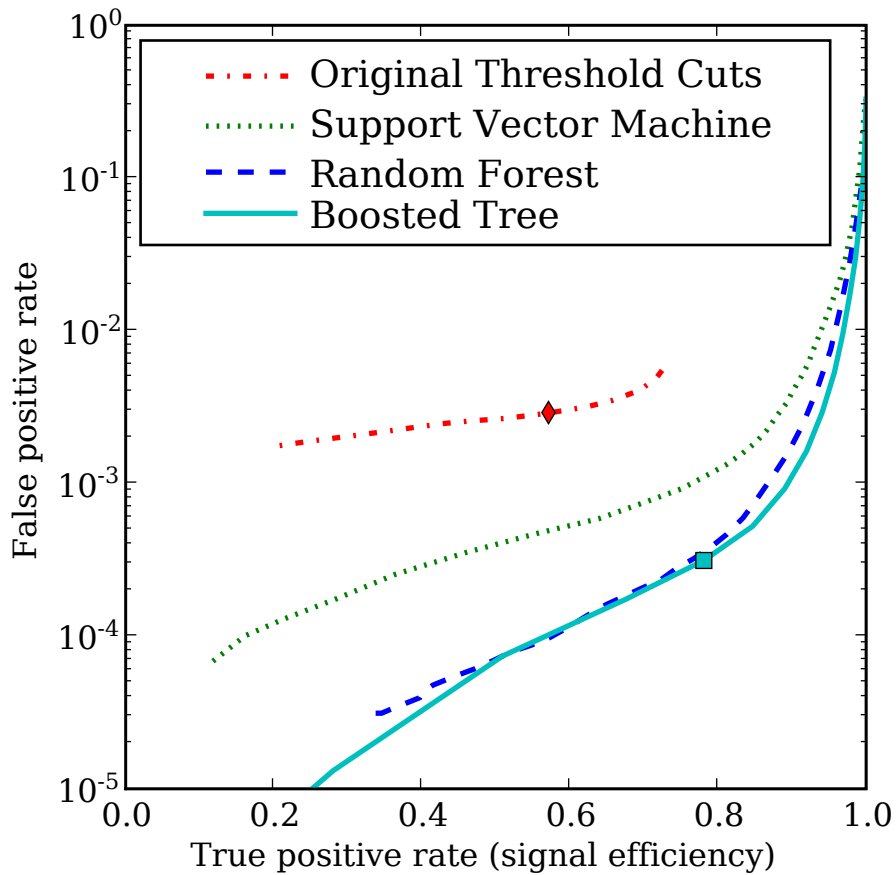}
\figcaption{Comparison of Boosted Trees (cyan solid line),
   Random Forest (blue dashed line), SVM (green dotted line),
   and threshold cuts (red dash-dotted line)
   for false positive identification fraction {\it vs.}~true
   positive identification
fraction.  For the threshold cuts, the signal-to-noise ratio,
motion, and shape cuts were varied to adjust signal and background rates.
The red diamond shows the performance of the threshold cuts used
during the SNfactory Summer 2006 search;
the cyan square shows the performance achieved with Boosted Trees
which were used for the Fall 2006 SNfactory search.
The lower right corner of the plot represents ideal performance.
\label{fig:eff}
}
\end{figure}

Figure \ref{fig:eff} shows the performance of several classification
methods applied to the SNfactory dataset.  The red square shows the
performance of the original threshold cuts upon which we were working
to improve.  The curves show that SVM, Random Forests, and Boosted Trees
all performed dramatically better than the threshold cuts
across a wide range of signal and background efficiencies.
The object features in our data have significant outliers which
prevented Fisher Discriminant Analysis
from being a useful classification method.

The overall best performance was obtained using Boosted Decision Trees,
with Random Forests providing nearly as good performance
with faster training and evaluation times.
Although SVM performed considerably better than threshold cuts, it
was not as successful as Random Forests or Boosted Trees.
A possible explanation is that the SVM proved to be more sensitive to
signal events that lie close to background events in the feature
space,
and could not strike a balance between modeling such
events {\it vs}.~overfitting to noise.
For example, a dim young supernova on a bright galaxy can be very similar
to a statistical fluctuation or a modest subtraction error in the images.
Robustness against overfitting is a known strength of boosted
classifiers \citep{bdt:freund99}, and the issue of overfitting
noisy data is an area of active research within the machine
learning community.
Further details of applying SVM to the
SNfactory dataset are described in \cite{svm:romano}.

Boosted Trees, Random Forests, and SVM successfully reduced the
false-positive rate for all types of background events in our data.
The three most common remaining background types are
faint optical ghosts from scattered light,
fluctuations in charge trails from bright stars due to CCD charge
transfer inefficiency,
and leftover dipoles from subtracting astrometrically misaligned objects.
The optical ghosts can be genuinely difficult to distinguish from
dim supernovae near our detection threshold.
The charge trails and dipoles are easy to distinguish by eye
but we currently do not have specific features which directly
address these two backgrounds, thus all classification methods
have difficulty with them, given the input features currently available.
These backgrounds are somewhat described by a roundness
feature and comparison of the flux in small {\it vs}.~large apertures.
Adding features to directly address these
backgrounds would improve the power of any classification method.
Projects with higher quality CCDs and optical designs to minimize
scattered light will naturally have fewer backgrounds as well.

The optimum selection criterion is dependent upon the tradeoff between
true positive selection efficiency (horizontal axis) and the false
positive selection efficiency (vertical axis).
At the SNfactory we seek
to maximize our signal efficiency within the realistic constraints of
the personnel and telescope time available to vet false positives.  For
the Fall 2006 search, we used Boosted Decision Trees and
choose a point with 10 times less background
than we had previously faced.  This corresponds to an average efficiency
of $\sim$78\% for identification of a supernova with a single filter
and one night of imaging.

\subsection{Optimizing Boosted Tree Parameters}
\label{sec:paramchoice}

Since Boosted Trees provided the best classification performance,
we describe here how the performance changed with various input
parameters to the Boosted Tree training.
The performance of Boosted Trees depends upon the number of trees generated
and the amount of branching which is done before finishing each tree.

\begin{figure}
\centering
\includegraphics{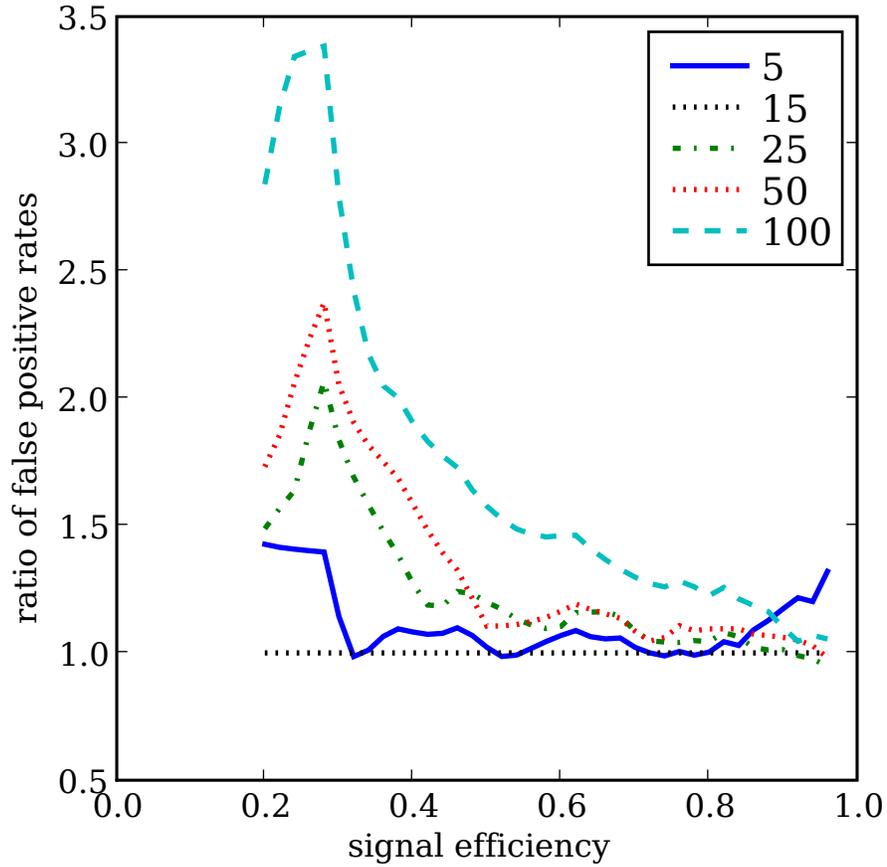}
\figcaption{
    A comparison of the performance of 200 boosted trees with varying
    leaf sizes.  10,000 training events were used; the plot shows the
    comparison of leaves with a minimum of $N=5,25,50,100$ events in comparison
    to the performance of the $N=15$ case.
\label{fig:nperleaf}
}
\end{figure}

For controlling the amount of branching per tree, the StatPatternRecognition
package has an adjustable limit on minimum number of events per leaf in
the final tree.
Our training
sample contained 5,000 signal and 5,000 background events.
We found the best performance with a minimum leaf size of 15 events,
which results in a set of boosted trees with 250--300 leaves each.
Figure \ref{fig:nperleaf} shows the relative performance of 200 trees with
a mininum leaf size of $n=5, 25, 50, 100$ relative to the $n=15$ case.

\begin{figure}
\centering
\includegraphics{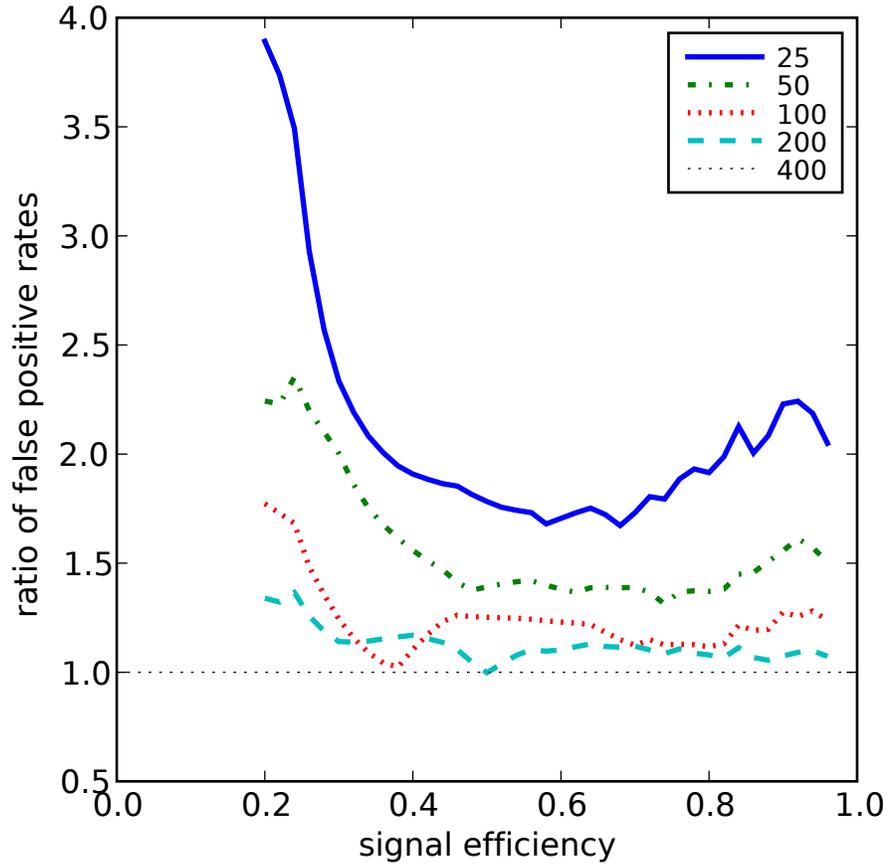}
\figcaption{
    A comparison of the performance of $N_{\rm tree}=25, 50, 100, 200$
    boosted trees
    with  a minimum of 50 events per leaf (out of 10,000 training events) in
    comparison to the $N_{\rm tree}=400$ case.
\label{fig:ntree}
}
\end{figure}

Figure \ref{fig:ntree} shows the relative performance of
$N = 25, 50, 100, 200$ trees in comparison to the $N=400$ case, using
a minimum leaf size of 50 events.
For the Fall 2006 SNfactory supernova search
we choose to use 200 trees with a minimum of 15 events per leaf out of the
10,000 training events.

\section{Combining Methods}

As expected, the various methods provided correlated output;
{\it i.e.}, events ranked highly by one classifier tended to be ranked
highly by another.  But even through Boosted Trees provided the best
classification performance overall, there were good signal events which
were found by SVM which were missed by the Boosted Trees
(for a given set of thresholds on the SVM and Boosted Tree outputs).
We attempted
to recover these events by combining the output of these two methods.
Several combinations were tried:
\begin{itemize}
\item Keep events which passed thresholds for either classifier.
\item Perform Fisher Discriminant Analysis on the output of the two
    classifiers.
\item Split the SVM {\it vs.}~Boosted Tree output space into sub-regions
    of signal and background.  This is conceptually similar to forming
    a decision tree in this 2D space and accounts for non-linear correlations
    in the SVM {\it vs.}~Boosted Tree output.
\item Use the output of SVM as an additional feature input for 
    Boosted Trees.
\end{itemize}
None of these methods produced results which outperformed the Boosted
Trees alone.  Although the combined classifiers could identify signal
events which would have been missed by just one classifier, the combination
also brought in an increased number of background events such that the
overall performance was the same or worse than the Boosted Trees alone.
This result is not generally true, however.  In other contexts, multiple
classifiers have been successfully combined to produce overall more
powerful results \citep{dietterich, kittler}.

\section{Discussion and Conclusion}

This work has shown a variety of object classification methods which provide
significantly better performance than is possible with the
method of threshold cuts used by most current supernova searches.
The implementations studied here used common defaults, such as
using the Gini parameter for optimizing Boosted Decision Trees
and the Gaussian kernel for SVM.  There are many variations of
these methods, some of which might provide further improved
performance.  But even these ``out of the box'' implementations
with minimal tuning provided much better performance than threshold cuts.

Any classifier will be limited by the quality and power of the input
features provided.  In practice,
after an initial round of training and validation,
one should study the misclassified events and introduce additional
features that distinguish these.  Such iterations are helpful
regardless of which classification method is used;
the main point of this work is to point to
new classification methods which will maximize the classification
power possible given a set of features.

As with any analysis, there is no substitute for clean data and a
well understood detector.  Problems which arise from false-positive
detections should first be addressed at the level of the detector
and data processing pipeline.  Future projects will hopefully have
the resources to address spurious detections at this level to make
the process easier for their object classifiers.
But even high quality, well
understood detectors and advanced image processing pipelines
such as SDSS will face signal {\it vs}.~background
classification problems, and this is where the methods described
in this paper come into play.

In addition to improved background rejection power, 
these new methods also have the advantage of generating
a single number which ranks the quality of an object rather than
a boolean pass/fail decision.  One may then adjust a threshold cut 
on that single number to tune the desired tradeoff between purity
and completeness.  Future surveys may publish transient alerts
using relatively loose quality requirements; subscribers to these
alerts can then place their own cuts on this quality rank to adjust
the purity, completeness, and input data rate as needed.

Boosted Trees, Random Forests, and Support Vector Machines all provide
much better object classification performance than traditional threshold
cuts.  When applied to the SNfactory supernova search pipeline, Boosted
Trees enabled us to find more supernovae with less work: Our efficiency
for finding real supernovae increased while our workload for scanning
non-supernova objects dramatically decreased.
Methods such as these will
be crucial for maintaining reasonable false positive rates at
the automated transient alert pipelines
of upcoming projects such as PanSTARRS and LSST.

\acknowledgements

We would like to thank
G.~Aldering,
S.~Bongard,
M.J.~Childress,
P.~Nugent,
and R.~Scalzo for useful conversations and 
assistance with scanning our supernova candidates.
We also thank the entire Nearby Supernova Factory collaboration for
confirmation and followup spectra of our selected candidates and
for the use of the search images for this study.
The anonymous referee provided many useful comments for which we
are grateful.

We are grateful to the technical and scientific staff of the Palomar
Oschin telescope, where our supernova search data are obtained.
The High Performance Wireless Research and
Education Network (HPWREN)\footnote{http://hpwren.ucsd.edu},
funded by the National Science Foundation grants 0087344 and 0426879,
has provided a consistently reliable
network for transferring our large amount of data from Mt. Palomar in
a timely manner.

This work was supported in part by the Director,
Office of Science, Office of High Energy and Nuclear Physics, of the
U.S. Department of Energy under Contract No. DE-FG02-92ER40704,
by a grant from the Gordon \& Betty Moore Foundation,
by National Science Foundation Grant Number AST-0407297.
This research used resources of the
National Energy Research Scientific Computing Center, which is supported
by the Office of Science of the U.S. Department of Energy under
Contract No. DE-AC02-05CH11231.

SB would especially like to thank the organizers and hosts of the
Statistical Inference Problems in High Energy Physics and Astronomy Workshop
held at the Banff International Research Station (BIRS), which is supported by
the U.S. National Science Foundation, the Natural Science and
Engineering Research Council of Canada, Alberta Innovation, and
Mexico's National Council for Science and Technology (CONACYT).

{\it Facilities:}
\facility{PO:1.2m (QUEST-II)}

\clearpage

\clearpage



\begin{figure}
\centering
\includegraphics{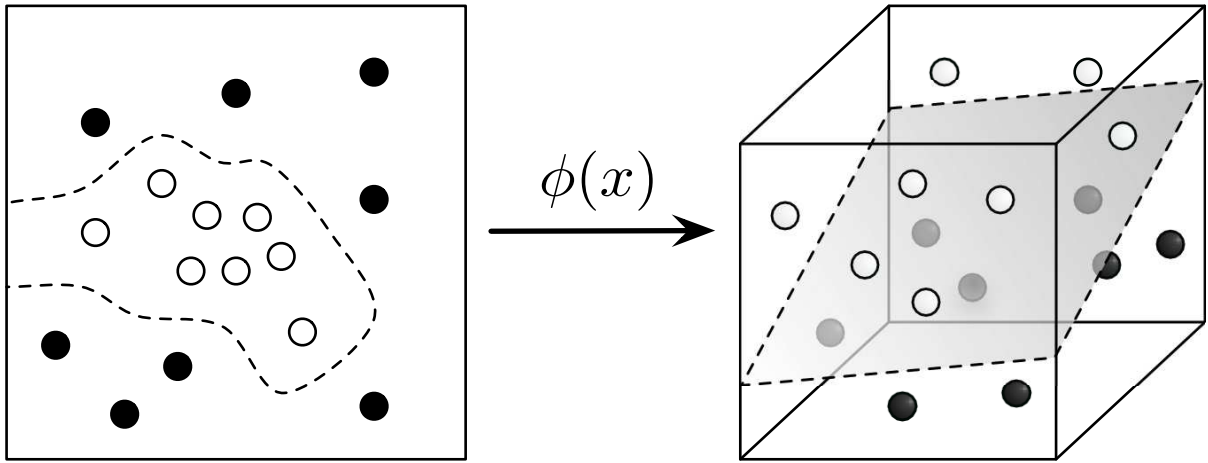}
\figcaption{
Support Vector Machines map an input space of features into a
    higher dimensional space where the separation of classes becomes
    easier.  The separation boundary in the original space
    may be quite complex, even disjoint.  In the higher dimensional
    space, the separation surface is a hyperplane whose parameters are
    entirely determined by the subset of events (the support vectors)
    nearest to the boundary.
\label{fig:ann}
}
\end{figure}

\clearpage

\clearpage

\clearpage

\clearpage

\end{document}